# The Inferred Cardiogenic Gene Regulatory Network in the Mammalian Heart

Jason N. Bazil[1,†], Karl D. Stamm[2,†], Xing Li[3], Timothy J. Nelson[4], Aoy Tomita-Mitchell[2] and Daniel A. Beard[1,*]

[1]Department of Molecular and Integrative Physiology, University of Michigan, Ann Arbor, MI
[2]Biotechnology and Bioengineering Center, Medical College of Wisconsin, Milwaukee, WI
[3]Division of Biomedical Statistics and Informatics, Department of Health Sciences Research, Mayo Clinic, Rochester, MN
[4]Departments of Medicine, Molecular Pharmacology and Experimental Therapeutics, and Mayo Clinic Center for Regenerative Medicine, Rochester, MN
[*]Correspondence: beardda@umich.edu (D.A.B.)
[†]Contributed equally to the manuscript

Running title: Inferring the Cardiogenic Gene Regulatory Network

Subject Category: Metabolic and regulatory networks / Simulation and data analysis

Character count: 31, 659

**ABSTRACT**

Cardiac development is complex, multiscale process encompassing cell fate adoption, differentiation and morphogenesis. To elucidate pathways underlying this process, a recently developed algorithm to reverse engineer gene regulatory networks was applied to time-course microarray data obtained from the developing mouse heart. The algorithm generates many different putative network topologies that are capable of explaining the experimental data via model simulation. To cull specious network interactions, thousands of topologies are merged and filtered to generate a scale-free, hierarchical network. The network is validated with known gene interactions and used to identify regulatory pathways critical to the developing mammalian heart. The predicted gene interactions are prioritized using semantic similarity and gene profile uniqueness metrics. Using these metrics, the network is expanded to include all known mouse genes to form the most likely cardiogenic gene regulatory network. The method outlined herein provides an informative approach to network inference and leads to clear testable hypotheses related to gene regulation.

# INTRODUCTION

Reverse engineering of a gene regulatory network (GRN) is an inverse problem that remains a significant challenge [1-5]. Despite high-throughput gene expression data obtained from methods such as real-time PCR [6], high-density DNA microarrays [7,8] and RNA Seq [9], complex interactions embedded in GRNs often overwhelm current methods of network inference [10,11]. Thus, there exists a need for new systematic tools to aid in the identification of the underlying architecture in networks like GRNs [12,13].

A general approach to reverse engineering of GRNs involves clustering genes into hierarchical functional units based on correlations in expression profiles [14]. To infer the causal relationships between these functional units, time-lagged correlation analysis is often employed [15,16]. Other identification methods include genetic algorithms [17], neural networks [18], and Bayesian models [19]. Several additional methods have been suggested to infer GRNs from expression data using prior knowledge of the GRN, perturbation responses, and other techniques (for details, see [4,10,20-23]). However, most of these methods rely on linear relationships to reconstruct the network without considering any combinatorial effects, noise or time delays. As a result, these approaches fail to capture the inherent nonlinearity of the interactions and interdependencies within the network [24]. To capture these complex, nonlinear interactions in gene expression patterns, general measures of dependency based on mutual information have been used [25-29]. However, mutual information does not give interaction directions and requires a significant amount of initial data which limits its use. To circumvent many of these issues, a new approach that relies on a combination of linear and nonlinear relationships to account for the dynamic nature of biology was developed [30]. Though the approach was validated with *in silico* data, the present study represents the first large-scale application to a dataset derived from a biological process such as cardiogenesis.

Cardiogenesis is the process in which the mesoderm of the embryonic blastocysts forms the fetal heart through a series of transformations (for review, see [31,32]). Morphology of heart development is well documented, but it is unclear how gene products regulate this process *in vivo*. With the advent of high-throughput technologies in genome-wide expression profiling to complement the single gene/pathway focus in the field of heart development, recent work has begun to address this complex transformation and identify key cardiopoietic factors that commit embryonic stems cells towards the cardiac-lineage genetic program [33,34]. Gene dosage drives protein expression and normal development as evidenced by myriad knockdown experiments. A survey of copy-number variable cardiac developmental genes has shown an enrichment of perturbed gene dosage in human children with congenital heart defects [35], emphasizing the role of molecular expression levels in dynamical networks. Going beyond curated candidate genes and identifying novel gene-gene interactions is an alternative and complimentary strategy to prioritize high value targets that may be overlooked with strategies relying on *a priori* annotations. Now the challenge is to determine how those key molecules come together in systems level analyses to create a fully functional organ.

Herein, we detail an approach to reverse engineer the cardiogenic gene regulatory network using a unique network inference algorithm. Time-course microarray data obtained from developing mouse hearts [36] were input into the inference algorithm to obtain cardiogenic gene regulatory networks. The approach is performed in two phases. First, a purely data-driven network inference algorithm is used on a subset of genes to construct an informative network that is then pruned to reveal the most likely GRN that best characterizes the input data. Second, this network is used as a scaffold to include additional genes from the entire dataset. A final filtering step using both Gene Ontology semantic similarity and expression profile clustering metrics yields a reduced network of higher confidence. This expanded network best characterizes the cardiogenic gene regulatory network as inferred by the algorithm.

# RESULTS

**Subnetwork Ensembles Predict Regulatory Interactions**

The gene profiles in the CG list were organized using hierarchical clustering. Figure 1A shows the clustered expression levels for the 171-gene CG for the nine time points from the beginning of development in the embryonic stem cell stage (R1) to the adult stage (A). The data show that the known pluripotent transcription factors (e.g. *Oct4*, *Nanog*, *Sox2* and *T*) peak at the early stages of heart development. During development, known cardiogenic genes activate (e.g. *Nkx2-5*, *Myl7*, *Notch1* and *Myog*). At the adult stage, ventricular cardiac specific markers (e.g. *Ttn*, *Myh6*, *Myh7* and *Ckm*) are significantly expressed. This comprehensive dataset provides a natural roadmap of the dynamic gene expression patterns that synchronize cardiac maturation and predicts genes previously unrecognized as cardiogenic contributors.

Typical examples of subnetwork models produced by the network inference algorithm that are capable of facilitating data-consistent simulations are shown in Figure 1B (left column). These results demonstrate a few interesting features of the subnetwork ensembles. First, all model trajectories pass through or near the experimental data. Without additional data, each model simulation presented is equally valid. Second, in some instances, the model simulations do not significantly vary. For example, the model simulations for *Myl2* do not diverge much from each other. Third, the predicted dynamics can vary considerably between time points as shown by the *Nppa* and *Aldh1a2* examples. This is due to uncertainty resulting from the model parameterization. This dynamical uncertainty introduces a unique opportunity for the design of optimal experiments (for details, see [37,38]). Typically each subnetwork set contains hundreds of data-consistent simulations, however some subnetwork ensembles contain only a few dozen that are capable of supporting data-consistent simulations, as shown by the Aldh1a2 example. In this case, the algorithm had trouble finding the right combination of regulatory interactions capable of fitting the data. There are only approximately 50 different model simulations presented for this example. While each subnetwork ensemble varies in the population size and dynamical uncertainty, they all support data-consistent simulations and thus all represent the possible regulatory interactions for the true CG GRN.

The dynamical uncertainty demonstrated in the *Nppa* and *Aldh1a2* examples is shown in their corresponding regulation frequency distributions. (Figure 1B, right). The height of each bar represents how often the network inference algorithm found that particular interaction sufficient to support data-consistent simulations. In these examples, many different regulatory interactions equally explain the data. The labeled stars signify regulatory interactions present in the IPA database. In approximately 11% of cases, the most significant regulator identified by the algorithm is reported in the IPA database. Among the population of all regulators identified by the algorithm approximately 3% are reported. In some cases, the highest ranked regulators returned by the algorithm are not previously reported and are thus targets for experimental validation. For *Nppa* and *Myl2*, the network inference algorithm identified a likely regulator, Gata4, along with many other regulatory interactions. Many of these additional interactions may be interpreted as being noise; however, among this noise are 14 (for *Nppa*) and 5 (for *Myl2*) regulatory interactions found in IPA. The graph for the *Aldh1a2* (bottom) shows a slightly different scenario. The number of interaction detected is lower than for *Nppa* and *Myl2*, and the algorithm did not detect any dominant regulatory interactions.

**Algorithm Performance Measures**

From the network inference algorithm, the likelihood that a given regulatory interaction is represented in the true CG GRN is assumed to be proportional to the frequency that that interaction appears in the subnetwork ensembles. Applying this confidence score to filter the recovered network edges, it is possible to explore the algorithm's performance measures when compared to the IPA network as shown in Figure 2 and Figure S1. Due to the incomplete nature of IPA, this approach to test the performance of the algorithm can be thought of as determining a lower bound of the algorithm's performance.

Figure 2A and B show the precision and recall curves as the confidence score threshold is increased. The precision starts relatively low at 0.03 and reaches a maximum of 0.12 at the cutoff value of 30%. Importantly, this indicates that strongly weighted edges are enriched with true positives. The recall decreases from a starting value of 0.8 as the cutoff value increases. This behavior is expected as the algorithm is designed to favor false positives over false negatives. The concept behind this design is that it is easier (from an experimental point of view) to remove a false positive via a simple perturbation experiment using the known topology versus searching in an unknown topological space for false negatives. Although most of the true positives appear to be of "low" importance based on the confidence score, all the precision scores are significant when compared to drawing edges from a random network (see Figure S1).

More typical performance measures reported with network inference algorithms are the receiver operating characteristic (ROC) curve and the precision recall (PR) curve [39]. The average performance for each measure is given by the area under the curve (AUC). In Figure 2C, the ROC curve is shown with an AUC score of 0.73. This indicates that the algorithm does an acceptable job when comparing the false positive rate versus the rate of identifying a negative as a false negative. In Figure 2D, the precision versus recall (PR) curve is shown with an AUC score of 0.035. Although, it is not possible to directly compare the algorithms performance metrics with other published algorithms, it is possible to make a rough comparison with the inference algorithms that participated in the DREAM5 challenge that used similar datasets [11]. The AUROC and AUPR scores are significantly better and near equal to, respectively, the scores published in the DREAM5 challenge for the *S. cerevisae* network [11]. It's important to note that the dataset used in the challenge contained a significant amount of more data than available in this study's mouse heart gestational time-course.

**Filtering Network Using Confidence Metrics Reveals Scale-free, Hierarchical Networks**
Constructing a network from all the regulatory interactions identified by the network inference algorithm generates a highly connected 'hairball' network that closely mimics an exponential network [40] as shown in Figure 3A. This network contains all 171 genes linked together with 12,157 edges (84% of all possible connections). The connectivity distribution follows a Poisson distribution with λ approximately 150. The clustering coefficient distribution is flat and independent of the node degree, $k$. More informative networks are deduced from this hairball by pruning away the low confidence edges. As the confidence score threshold is increased, a more familiar topology is revealed, as shown in Figure 3B and C. With a moderate threshold enforced (3B, 15%), the connectivity distribution follows a power law distribution with γ equal to 1.19. The clustering coefficient distribution is a function of the inverse of the node degree with an $R^2$ value of 0.81. With a stricter threshold placed on the confidence score (3C, 30%), the connectivity distribution still follows a power law distribution, but γ increases to 1.31. Similarly, the clustering coefficient distribution remains a function of the inverse of the node degree, except that the $R^2$ improves to 0.90. Thus, as the confidence score cutoff is increased, the inferred networks possess a scale-free topology and hierarchical structure [40]. If the threshold is set too high (3D, 40%), the network becomes disjoint and connectivity metrics are inapplicable.

The network presented in Figure 3C is the most likely GRN that best characterizes the input data according to our network inference algorithm. This network contains four major hub genes (defined as genes having greater than six outgoing regulatory connections): *Gata4*, *Mapk1*, *Sox18* and *Whsc1*. *Gata4* has the most outgoing connections and is already a known hub gene [41] that is involved in a wide variety of processes involving embryogenesis, cardiogenesis, and muscle development; *Mapk1* is involved with embryo and cardiovascular development [42]; *Sox18* regulates processes such as embryogenesis, vasculogenesis, and cardiomyocyte differentiation [43]; and *Whsc1* is heavily involved in heart morphogenesis, particularly heart chamber formation and development [44]. Thus, the hub genes identified here are corroborated by their expected roles during well-established pathways critical for proper cardiogenesis.

**Most Confident Regulatory Connections are Enriched with Known Interactions**
Many of the genes directly downstream from the largest hub gene, *Gata4*, are in the IPA database. Moreover, a few genes further downstream that are recovered from the algorithm are also in the IPA database. These include *Nkx2-5* and *Mapk1*. The genes connecting *Gata4* and *Mapk1* are *Dppa5a*, *Ankrd1* and *Ttn*. All these genes share various cardiogenic annotations except for *Dppa5a*. This is either a novel finding or another gene with very similar expression to *Dppa5a* may be the actual gene that participates in this pathway. It turns out that *Dppa5a* expression is highly correlated with many other potential genes including *Tdgf1* and *Klf5*. These two specific genes share many of the cardiogenic annotations with the other genes in the pathway and may serve as this link between *Gata4* and *Mapk1*. Experimental validation is ultimately required to confirm the role of *Dppa5a*, *Tdgf1* and *Klf5* in this pathway. This example highlights the difficulty in identifying precise topologies based solely on expression profiles.

The top 25 gene interactions identified by the network inference algorithm based on their fidelity scores are shown in Table I. The fidelity scores fall on a normal distribution with a heavy right tail having a mean of 0.98 and a standard deviation of 0.35. Most of the results consist of novel interactions; however, 30% of them are reported in the IPA database. Those edges with $Z_k>0.21$ are enriched with interactions annotated by IPA, hypergeometric p-value < 0.05 as shown in Figure S2. Peak significance (p=0.0006) is achieved at $Z_k>1.885$ although this metric becomes noisy as fewer interactions are considered. As with the confidence metric cutoff, pruning the network using the fidelity scores produces scale-free, hierarchical networks.

Among the list of genes in Table I and Supplementary Table 1, *Fli1* is the most promising candidate gene predicted to be involved in cardiogenesis. This gene encodes a transcription factor containing an ETS DNA-binding domain and may be involved with a variety of biological processes such as cellular differentiation, proliferation, migration, apoptosis and angiogenesis [45,46]. Although there is no direct evidence of its role in cardiogenesis, it is essential for embryogenesis and endothelial gene expression [47]. Furthermore, *Fli1* increased expression has been linked to decreased cardiac fibrosis in a physiological model system of cardiac damage and may imply a regulatory role not previously recognized [48]. Another interesting prediction in Supplementary Table 1 is *Tbx18* inhibiting *Sox7*; both of these genes are important early transcription factors. *Tbx18* has been shown to convert cardiomyocytes into pacemaker cells, and plays a role in tissue engineering [49]. This could be used to model cardiac pathologies such as atrial arrhythmias or ventricular arrhythmias [50] and provides *in silico* prioritization of gene therapies.

Supplementary Table 1 consists of many well-characterized interactions that are important for cardiogenesis. Among them is a particularly important interaction involving *Nkx2-5* activation by *Gata4*. Durocher et al. demonstrate *Gata4* binds to the C-terminus autorepresisve domain of *Nkx2-5* and activates this transcription factor [51]. Another interaction identified in Supplementary Table 1 is between *Sox18* and *Sox7*. These two transcription factors have been described to act concomitantly during cardiac and vascular development [52,53] which suggests the existence of a mutual feedback type of regulation. Also, *Fog2* (*Zfpm2*), a cofactor of *Gata4*, is recognized as an inhibitor of *Gata4* activity [54,55], although not necessarily of *Gata4* expression. The functional result of this interaction was also identified by the predictive network as *Zfpm2*-mediated inhibition of *Gata4*. Finally, some directionally undefined interactions such as that between *Tbx20* and *Gata4* [56,57] are resolved in the networks, in this case pointing at a *Tbx20* activation by *Gata4*.

**Network Expansion Reveals Novel Regulatory Modes**
Although the input data (CG list) consist of less than 200 cardiac related genes, Figure 1A shows that the expression of early transcription factors in the embryonic stem cell stage and heart tube activates a wave of gene expression that leads to the sustained expression of adult cardiomyocyte related genes. Assuming this dataset captures this phenomenon reasonably well, it is possible to reach into the entire mouse dataset [36] and identify a representative cardiogenic GRN in the mammalian heart. The initial network returned by the algorithm was used as a scaffold and the entire mouse heart dataset (consisting of more than 20,000 genes) and utilized to expand the network. The expanded network consists of 1,080 genes and 63,558 edges as given in Supplementary Table 2.

The most significant edges of this network are shown in Figure 4. This network is filtered down to 524 genes and 1,750 edges by removing edges with a low confidence and fidelity score. It best characterizes the cardiogenic gene regulatory network as inferred by the network inference algorithm. Gene annotations reveal a majority of genes are involved with embryogensis, the development of the cardiovascular system, heart morphogenesis, muscle energetics and epigenetics. The p-values for all the go terms selected using ClueGO in Supplementary Table 3. Approximately 391 of the genes had representative GO annotations.

Over half of the representative GO terms for this network are Cardiovascular System Development and Embryo Development. There is a heavy cluster of genes with these annotations that consist of many transcription factors (e.g. the Fox, Gata, Tbx, Sox and Zic families) among sparsely interwoven genes involved in cell signaling, cell migration and metabolism. This cluster serves as the central network hub that connects the rest of the network and coordinates gene expression for a variety of biological processes. For example, there is a small cluster of cell adhesion related genes (e.g. the Hapln family along with *Ntm* and *Pcdh7*) near the top-center of the network shown in Figure 4. This cluster appears to be under the control of *Lama4*, a protein involved with cell adhesion, differentiation, migration and signaling. *Lama4* links these cell adhesion genes through a pathway involving the extracellular matrix proteins *Dpt* and *Col13a1* which are connected to *Lbh*, a major hub gene located in the central network hub cluster.

In another example, a cluster of genes involved in metabolic processes is found in the bottom-left part of Figure 4. These genes mainly code for proteins involved with CoA-mediated metabolic processes. Just above them, is another cluster of genes that do not have representative GO terms. The ClueGO analysis did not assign any; however, many genes in this cluster are involved with mitochondrial energetics (e.g. *Cox8b* and *Cox6a2*), redox-mediated signaling (e.g. Dhrs family) and other metabolic processes (e.g. *Dlat*, *Ldhd* and *Acad10*). These two clusters are linked via *Hadha*, a protein that catalyzes the last steps of mitochondrial beta-oxidation. ClueGO analysis indicates that *Hadha* is also involved with signaling and may serve as a putative link between these two clusters. These two clusters are connected to the central network hub via *Alkbh8*, a gene that may be important for angiogenesis [58].

## DISCUSSION

The approach presented herein relies on a purely data-driven inference algorithm coupled to an informative association and filtering method. In doing so, the most likely predictable gene interactions obtained from the algorithm are those appear often in the subnetwork ensembles, whose genes cluster infrequently with other genes and those that share many GO terms. Roughly 30% of the top 25 gene pairs identified using the CG list are previously known. This is a significant achievement considering that the approach is data-driven, supplemented with an ontology library and relies on a computational model to approximate gene expression. The network was expanded using the same semantic similarity and gene profile metrics to include all genes in the mouse genome to form the most likely cardiogenic gene regulatory network as predicted by the algorithm. To further validate the inference approach, the most significant regulatory interactions will be tested using induced pluripotent stem cells driven towards cardiomyogenesis.

Methods used to construct networks using gene expression profiles are typically undermined by the similarity of the expression between various genes in the dataset. This makes assigning network edges challenging since a given regulatory interaction can be also explained by swapping out the source gene with another gene that has a very similar expression profile. Genes of this nature have been called module genes [59]. A mitigating strategy is to focus on interactions that can be explained by relatively few genes and share common pathways. This is done by filtering out gene interaction pairs whose cluster product is large and whose annotations widely differ. This is achieved by keeping only the gene pair interactions with the highest fidelity scores (Equation 1). Gene interaction pairs with small cluster products and high semantic similarity scores are more likely to be previously identified in the IPA database as shown in Figure S1. This type of approach has recently been utilized to construct gene networks and shown to produce superior results when compared to more traditional methods [60].

While the algorithm is among the most efficient in its class, it is still computationally expensive to exhaustively search all possible combinations of gene interactions. In order to reduce the complexity of the inferred networks, the algorithm can be extended to exploit additional information obtained from pathway analyses and other sources of independent data. This will lead to an algorithm that produces more experimentally testable hypotheses, result in more efficient network inference and deliver more relevant biological networks. In addition, the approach presented herein is well suited to increase our collective understanding of the processes involved with cell lineage commitment, characterize the progression of polygenic diseases and help unravel the complexities associated with pharmacogenomics.

## MATERIALS AND METHODS

### Cardiogenesis Data

The time-course gene expression data of the mouse heart were obtained at sequential heart development stages using Affymetrix Mouse Genome 430 2.0 microarrays [36]. Gene expression data were obtained using the RMA algorithm [61] at nine developmental stages consisting of embryonic stem cells (ESCs as starting point), early and late embryonic stages until the adult stage. At embryonic day nine (E9.5) and later, the left and right ventricles were separated, and gene expression was assayed for each chamber. This was done for all time points after E9.5. The microarray platform used in this time-course data contains 45,000 probesets representing known genes in the mouse genome. This large number prohibits the use of methods of inference that rely on model simulation with current computational capabilities and algorithms. Therefore, in order to complete the network inference in a reasonable amount of time, a manageable list of genes of interest was generated from the entire mouse genome expression data based on the following selection criteria: i) the top 50 differentially expressed genes ii) the top 50 differentially expressed transcription factors and iii) a list of cardiac specific genes that are believed to be involved with a variety of congenital heart diseases [62]. The final list consisted of 171 genes (herein, the cardiogenesis list, or CG list), and the corresponding expression profiles are representative of the left ventricle of the heart. The expression data for these genes are used in the initial inference which is detailed below. Figure 1 shows a heat map of the expression for the CG list across all nine time points. For modeling purposes, the nonnegative RMA-normalized data is scaled between zero and one.

**Network Inference**

Our previously described algorithm [30] was used to model the expression levels of the genes in the CG list during the development of the heart. In brief, the network inference algorithm splits an N-dimensional problem into N 1-dimensional problems, one for each observed state variable (gene). Putative regulatory networks associated with each of the individual state variables are independently identified. Network identification for each variable/gene is based on a generalized model of gene expression dynamics accounting for competition between activation and inhibition from all other genes in the dataset. Since this problem is typically under-determined, ensembles of putative subnetworks are developed for each gene. A subnetwork is a type of subgraph that only contains the target node and the neighboring regulatory nodes. Each ensemble contains anywhere from 50 to 2,000 subnetworks that support data-consistent simulations. A data-consistent simulation is defined as one that leads to a variance-weighted least squares error function less than 0.75. (See Figure 1B for examples.) Putative regulatory networks for the full 171-gene list were generated by randomly sampling and combining the subnetworks. In total, 1,000 putative networks were generated and statistical information of the gene interaction pairs, or edges, regarding frequency of occurrence, directionality, and regulatory strength (activating versus inhibiting) was collected. The full set of statistical metrics on all gene pairs is given in Supplementary Tables 1 and 2. Subnetwork generation and network analysis was done using MATLAB R2011b (The Mathworks, Inc.). Network analysis was done using toolbox published by MIT's Strategic Engineering Research Group [63]. Network visualization employed Cytoscape 3.0.2 [64]. Gene ontology (GO) annotations for mouse were obtained and analyzed using the GO biological process terms (2/07/13 release) using ClueGO [65] with the kappa score threshold set to 5, GO term fusion turned on, and the rest of the options at the default settings.

**IPA Validation**

To benchmark the network inference algorithm, a database of accepted gene regulatory interactions is required. Ingenuity Pathway Analysis (IPA) was utilized (Ingenuity® Systems, www.ingenuity.com) as an expert-curated gene interaction database that is regularly updated and maintained. While the regulatory interactions in IPA are not complete, it provides a curated database to verify results from our inferred networks. It is important to note that a regulatory interaction *cannot* be ruled out because it does not appear in the IPA database since many interactions have yet to be discovered and documented. Exported regulatory interactions from IPA lack directionality, so analysis was done treating the network as an undirected graph. Network enrichment calculated as hypergeometric with R *phyper*. The interaction list for each gene in the CG list was downloaded from the IPA servers no later than 07/24/2012.

**Gene Ontology Semantic Similarity**

A scoring of each predicted gene-gene interaction is computed to enable pruning our predictions by estimating the pair's biological relevance. Semantic similarity scores were calculated using pre-propagated GO biological Process terms for the mouse genome obtained from Gemma [66] using the method described in Mistry and Pavlidis [67]. Gene Ontology is a hierarchically structured controlled vocabulary, and most genes have multiple GO annotations. The pre-propagated annotations give a single path from the GO Biological Process root to the gene's most specific leaf node allowing the exact term set to become a proxy for the gene's biological role. A GO term set-intersection for any pair of genes quickly yields a biological similarity metric (between 0 and 100%). These are used in a late stage filtering to choose those pairs believed to share a common biological process.

**Expression Profile Clustering**

As the network inference uses only numerical profiles and is agnostic to each gene's true identity, common profiles will confound results. Stochastic clustering using self-organizing maps (SOM) is used to cluster gene profiles [68,69]. As the clustering is sensitive to initial parameters, many iterations are performed and a count of gene-gene co-clustering is collected. The SOM forces all genes into a grid layout, varied randomly in size from 3x3 through 50x50 to achieve a balance of precision and smoothing. A total of 10,008 genes with stable, non-dynamic expression (defined as dispersion, or standard deviation over the mean less than 20%) are considered too common and are excluded from the SOM evaluation, leaving 11,307 genes. A total of 2,240 SOMs were computed and pooled together to determine a given pair's coincidence frequency. Gene profiles with similar time-course dynamics will often cluster together and have high co-incidence scores. A threshold of 70% was used to partition the 21,315 gene set, resulting in 4,099 clusters. The number of genes in each cluster is approximately exponentially distributed.

**Fidelity Score**

A metric to gauge the fidelity of a predicted interaction is constructed to maximize biological relevance while minimizing gene profile ambiguity (Equation 1). The GO term overlap semantic similarity scores are represented by the Jaccard index [70], z-transformed. The product of the cluster sizes of the $i$th and $j$th gene in pair $k$, where $k \in \{i, j\}$, are computed using the clusters identified with the SOM method. Both the Jaccard indices and cluster products are orthogonal metrics and log-normally distributed. As such, the fidelity score for $k$th gene pair, $Z_k$, is the difference of the z-score of the log of the Jaccard index, $Zj_k$, and the z-score of the log of the cluster products, $Zc_k$. By maximizing $Z_k$, the $k$th gene pair shares a high degree of semantic similarity and infrequently clusters with other genes. In other words, gene pairs that possess large fidelity scores are the most relevant predictions.

$$Z_k = Zj_k - Zc_k \qquad \text{Equation 1}$$

**The Cardiogenic Network**

The network constructed from the initial dataset was expanded to include additional genes that were not originally included in the input data. This was done by first constructing an eigengene network analogous to the method by Langfelder and Horvath [59]. An eigengene network consists of a network of unique gene modules that best characterizes the network in a reduced form. A gene module consists of a set of genes with highly correlated expression profiles. Our gene modules are derived from the clusters inferred by the SOM coincidence. Each predicted interaction is expanded to include all possible combinations of gene pairs. For example, if gene A is predicted to interact with gene B, but gene A has four additional genes in its profile cluster and gene B has six others, a total of five times seven gene pairs could be represented by the numerical prediction. This expanded set is ranked and filtered by Equation 1.

**ACKNOWLEDGEMENTS**

We thank Almudena Martinez-Fernandez for her comments and help with the manuscript. This work was supported by grant P50-GM09450 from the National Institutes of Health.

Figure 1. Overview of experimental data and modeling results. (A) Hierarchical clustering of the mouse heart gene expression input dataset. After E8, the data is representative of gene expression in the left ventricle. The cardiogenic program is seen to propagate through the network yielding elevated expression of the typical cardiomyocyte markers by the Adult stage. The CG list profiles were clustered using the MATLAB clustering algorithm using the Pearson correlation and complete linkage metrics. (B) Example results from the subnetwork analysis are shown. The algorithm returns data-consistent, simulated gene expression profiles that often show some degree of dynamical uncertainty between the data. Each line represents a separate model simulation that may or may not have the same network topology. The corresponding regulation histograms demonstrate another property of the algorithm. The height of each bar represents the fraction a given regulator appears in a given subnetwork ensemble. Therefore, it reflects a measure of confidence or significance for that particular gene interaction. Stars labeled with a gene name signify IPA validated edges.

Figure 2. Performance metrics of the algorithm using a network created using the IPA database as the "gold standard." The confidence cutoff metric is defined as the fraction a given edge appears in the network as described in the legend of Figure 1. (A) As the lower confidence edges are pruned from the network, the precision of the algorithm increases which indicates that edges that often appear in the subnetwork ensembles are enriched with validated interactions. (B) Conversely, the recall of the algorithm decreases as the precision increases which indicates many validated interactions are of low confidence. (C) The area under the curve (AUC) for the receiver operating characteristic demonstrates that the algorithm does an acceptable job of identifying validated interactions relative to missing them. The diagonal line corresponds to random prediction. (D) The AUC for the precision recall curve demonstrates the algorithm is equally good if not better when compared to other algorithms that explored similar types of data [11]. The horizontal line represents random prediction. See also Supplementary Figure 1.

Figure 3. Predicted gene regulatory network of 171 nodes at edge weight cutoffs 0, 15, 30, and 40% (Panels A-D respectively). The complete network (A) shows the 'hairball' characteristic of an exponential network (node connectivities follow a Poisson distribution with $\lambda \approx 150$, and clustering coefficient distribution is independent of the node degree). Interactions also found by *Ingenuity Pathway Analysis* are marked with stars. Edge thickness represents the occurrence frequency or weight. Edge color is red for inhibiting, green for activating, and yellow for unclear relationships. Edge arrowheads are marked for interactions with clear directionality. As the cutoff metric is raised, scale-free, hierarchical networks emerge. At a cutoff of 15% (B), the connectivity distribution follows a power law with $\gamma=1.19$, and the clustering coefficients scale with the reciprocal of the node degree with $R^2=0.81$. At a cutoff value of 30% (C), the network further represents a scale-free, hierarchical network where $\gamma=1.31$ and $R^2=0.90$. At 40% (D) only a few strongly predicted interactions remain, $\gamma=1.70$ and the network is too sparse to compute the clustering coefficients. See also Supplementary Table 1 and Supplementary Figure 2.

Figure 4. The inferred network using the CG list was used as a scaffold and extended to include genes from the entire mouse genome by expression profile similarity. Representative annotations using the Gene Ontology database are shown by node color key. All annotations are relevant to cardiogenesis with some more specific than others. Edge color and thickness are as in Figure 3. Directional arrows are not shown for clarity. The gene interactions shown are the edges with the top 5% fidelity scores and the top 50% confidence scores. GO term acronyms: CM, Cell Migration; CSD, Cardiovascular System Development; CA, Cell Adhesion; CO, Cytoskeletal Organization; ED, Embryo Development; EMO, Extracellular Matrix Organization; MAMP, Monocarboxylic Acid Metabolic Process; RCC, Regulation of Cell Communication; SCD, Stem Cell Differentiation; TDD, Transcription, DNA-dependent; TRP, Transmembrane Receptor Protein Serine/Threonine Kinase Signaling Pathway. See also Supplementary Tables 2 and 3.

Table I. Top 25 Gene Interactions from the CG List

| Gene Pair Interaction | Confidence Score[a] | Fidelity Score | Found In: |
|---|---|---|---|
| Fli1 is strongly regulated by Sox7 via activation | 0.463 | 3.49 | - |
| Foxh1 is strongly regulated by Gata4 via inhibition | 0.389 | 3.17 | - |
| Chd7 is strongly regulated by Whsc1 via activation | 0.395 | 2.26 | - |
| Foxa3 strongly regulates Foxh1 via activation | 0.401 | 2.22 | - |
| Sox18 interacts with Sox7 via activation | 0.573 | 2.08 | - |
| Hba-a1 /// Hba-a2 strongly regulates Hbb-b2 via activation | 0.475 | 1.84 | IPA, R |
| Gata4 interacts with Sox18 via activation | 0.46 | 1.59 | - |
| Foxa1 is strongly regulated by Gata4 via inhibition | 0.37 | 1.49 | IPA |
| Gata4 strongly regulates Sall4 via inhibition | 0.422 | 1.48 | - |
| Ehmt1 regulates Sgcg via inhibition | 0.483 | 1.41 | - |
| Gpihbp1 is strongly regulated by Whsc1 via inhibition | 0.437 | 1.41 | - |
| Bcl11a strongly regulates Sema5a via activation | 0.468 | 1.28 | - |
| Gata4 strongly regulates Pou5f1 via inhibition | 0.362 | 1.20 | IPA |
| Gata4 strongly regulates Nppa via activation | 0.463 | 1.17 | IPA |
| Gata4 strongly regulates T via inhibition | 0.363 | 1.12 | R |
| Nfia is strongly regulated by Sox18 via activation | 0.389 | 1.12 | - |
| Hey2 strongly regulates Tnni1 via activation | 0.528 | 1.01 | - |
| Casq2 is strongly regulated by Whsc1 via inhibition | 0.464 | 0.97 | - |
| Gata4 strongly regulates Otx2 via inhibition | 0.42 | 0.87 | - |
| Ehmt1 strongly regulates Myom2 via inhibition | 0.507 | 0.83 | - |
| Gata4 regulates Nanog via inhibition | 0.452 | 0.65 | IPA |
| Hba-x is strongly regulated by Meox2 via inhibition | 0.365 | 0.64 | - |
| Eomes is strongly regulated by Gata4 via inhibition | 0.414 | 0.61 | - |
| Gata4 strongly regulates Zic3 via inhibition | 0.366 | 0.60 | IPA |
| Eln interacts with Kcne1 via activation | 0.782 | 0.54 | - |

[a]The fraction a given edge appears in the 1000 networks sampled from the subnetwork ensembles.
[b]Pair interaction found in IPA or Reactome (R) databases.

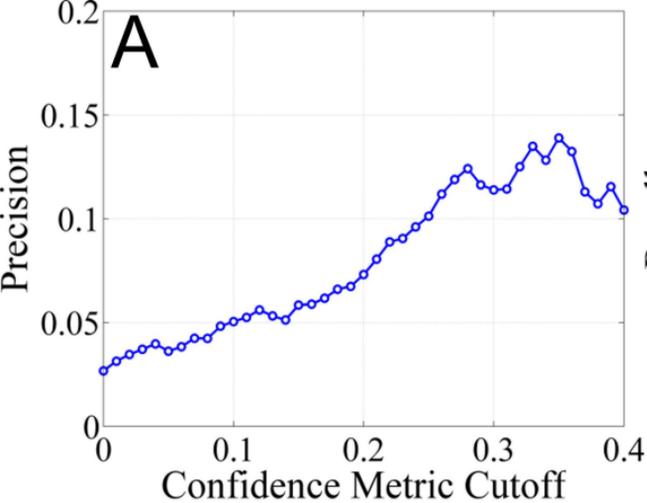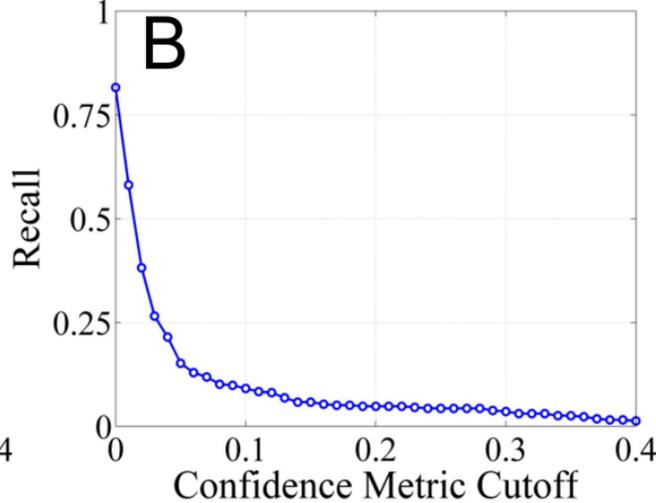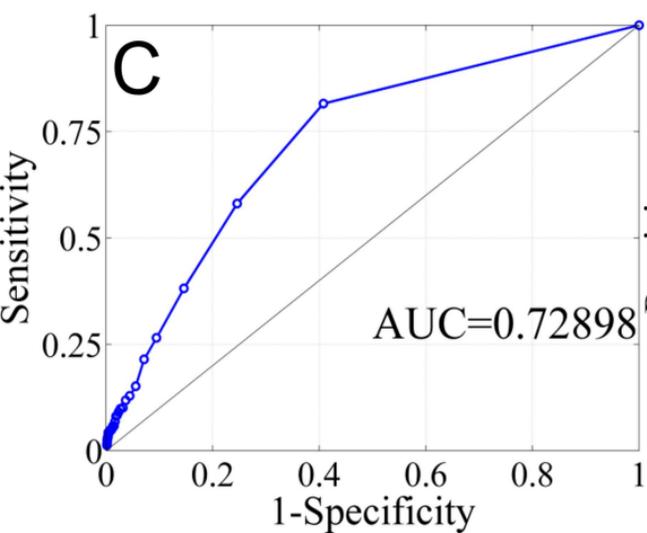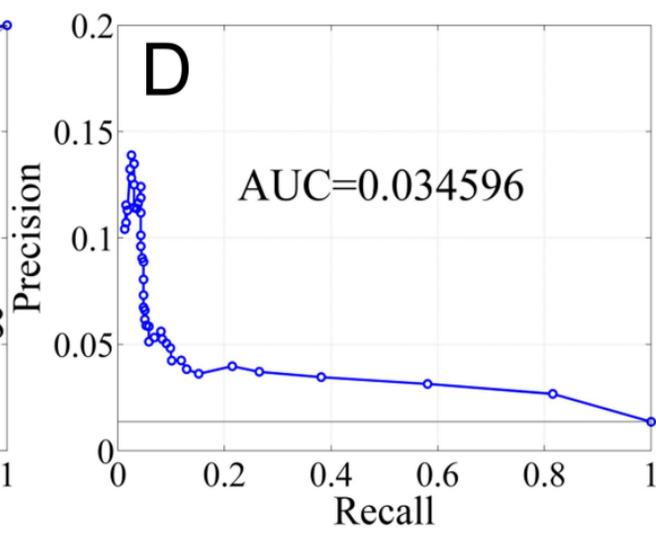

**Color Legend**
- CM
- CSD
- CA
- CO
- ED
- EMO
- MAMP
- RCC
- SCD
- TDD
- TRP
- Other